\documentclass[12pt, journal, onecolumn]{IEEEtran} 

\usepackage{booktabs}
\usepackage{multirow} 
\usepackage{cite}
\usepackage[cmex10]{amsmath}
\usepackage{amssymb} 
\usepackage{amsfonts}
\usepackage{amsthm} 
\usepackage{graphicx,epsfig} 
\usepackage{setspace}
\usepackage{algorithm}
\usepackage{algorithmicx}
\usepackage{algpseudocode}
\usepackage{enumerate}
\usepackage[margin=1 in]{geometry}

\newtheorem{Lem}{Lemma}

\newtheorem{Col}{Corollary}
\newtheorem{Conj}{Conjecture}
 
\usepackage{hyperref}

\begin{document}

\title{User Arrival in MIMO Interference Alignment Networks}
\author{Behrang~Nosrat-Makouei,~\IEEEmembership{Student Member,~IEEE,}
        Jeffrey~G.~Andrews,~\IEEEmembership{Senior Member,~IEEE,}
        and Robert~W.~Heath,~Jr.,~\IEEEmembership{Fellow,~IEEE}
\thanks{The authors are with the Department of Electrical and Computer Engineering, The University of Texas at Austin, Austin, TX 78712 USA (e-mail:
behrang.n.m@mail.utexas.edu; jandrews@ece.utexas.edu; rheath@ece.utexas.edu).}
\thanks{This work was supported by the DARPA IT-MANET program, Grant W911NF-07-1-0028, and by the Army Research Labs, Grant W911NF1010420.}
\thanks{This work was presented in part at IEEE Int. Conf. Acoust. Spch. Signal Process., Prague, Czech Republic, May 2011 \cite{BehrangICASSP2011}.}}
\maketitle
\begin{abstract}
In this paper we analyze a constant multiple-input multiple-output interference channel where a set of active users are cooperating through interference alignment while a set of secondary users desire access to the channel. We derive the minimum number of secondary transmit antennas required so that a secondary user can use the channel without affecting the sum rate of the active users, under a zero-forcing equalization assumption. When the secondary users have enough antennas, we derive several secondary user precoders that approximately maximize the secondary users' sum rate without changing the sum rate of the active users. When the secondary users do not have enough antennas, we perform numerical optimization to find secondary user precoders that cause minimum degradation to the sum rate of the active users. Through simulations, we confirm that i) with enough antennas at the secondary users, gains equivalent to the case of all the users cooperating through interference alignment is obtainable, and ii) when the secondary users do not have enough antennas, large rate losses at the active users can be avoided.
\end{abstract}
\section{Introduction}

\indent 
Interference alignment (IA) \cite{Cadambe2008a,Maddah-Ali2008} is an interference management technique broadly described as confining the interference to a subspace of the received signal space such that an interference free subspace becomes available for the desired signal. Unlike techniques such as ignoring the interference \cite{Etkin2008}, decoding/cancelling the interference \cite{Han1981} or avoiding the interference (orthogonal access), IA achieves the maximum degrees of freedom of the K-user interference channel. 

\indent Interference can be aligned using signal levels \cite{Sridharan2008}, time or frequency channel extensions \cite{Cadambe2008a}, and multiple antennas at the interfering nodes \cite{Yetis2010, Tresch2009a, Peters2011}. IA using the signal levels (lattice alignment) requires a baseband module not compatible with near future wireless networks \cite{ArunabhaGhosh2010}. IA over time extensions of the channel is a mere theoretical tool requiring non-causal knowledge of channel state information. IA over frequency extensions of the channel, in its original form \cite{Cadambe2008a}, requires a large number of channel uses to fully exploit the power of IA, and is not suitable when global frequency synchronization is not available.  Therefore, aligning the interference over the spatial dimension, i.e. IA over constant multiple-input multiple-output (MIMO) channels, is the most attractive physical layer signaling approach. In this paper, we focus on MIMO IA.

\indent In general, a high computational \cite{Gomadam2008,Papailiopoulos} and overhead \cite{Krishnamachari2009,Omar2010_AnalogConf} cost is associated with finding the IA precoders and combiners. Therefore, once a set of users have aligned their interference, it is desirable to retain their alignment status until the channels change. Most future interference-limited wireless networks, however, will be packet-switched with bursty data traffic, requiring frequent changes in the number of active users \cite{ArunabhaGhosh2010}. Also, the feedback overhead \cite{Steve2010_IAoverhead} and the number of antennas at each node \cite{Yetis2010} practically limit the number of user pairs that can cooperate through IA. Therefore, in an IA network, there will be nodes that are \textit{arriving} to the network and, although not included in the existing IA setup, wish to communicate with their receivers. Determining when such nodes can be admitted to the network (are allowed to transmit) and/or developing transmission strategies for them is referred to as \textit{user arrival}. Although there exist a few prior strategies for user arrival and precoder design for MIMO networks, they either do not consider IA or they are restricted to opportunistic 
access methods  \cite{Matskani2008, Butussi2006, Bartolome2006, Perlaza2008}.

\indent In this paper, we consider a set of active users exhausting the network resources by cooperatively utilizing IA and a further set of secondary users who wish to communicate in this network.
We assume the secondary users are required to have minimum impact on the performance of the active users, defined as the sum rate of the active users after a zero-forcing (ZF) receiver, and the active users ignore the presence of the secondary users when designing their precoders and/or equalizers. 
We first compute a \textit{zero-impact} threshold for the number of secondary transmitter antennas where secondary transmitters with more (or equal) antennas than this threshold can use the communications medium without degrading the sum rate of the active users. Then we find optimum and suboptimum secondary user precoders for two cases: (i) when the secondary users satisfy the \textit{zero-impact} threshold and (ii) when they do not.

\indent When the secondary users satisfy the \textit{zero-impact} threshold, there exists a set of precoders that will \textit{not} degrade the sum rate of the active users. Thus, the secondary users can optimize an objective function of their own link by selecting a precoder from this set. We choose the achievable sum rate as the objective function and derive optimum and suboptimum precoders for the case of one and two secondary users respectively. For more than two secondary users, we choose the degrees of freedom (DOF) as the objective function, defined as the slope of average sum rate (b/s/Hz) versus logarithm of signal-to-noise ratio (dB) at high transmit power. We propose \textit{successive IA} precoding and show it is optimum for various network setups determined by the number of active users, secondary users, and antennas at each node. When the secondary users have fewer antennas than the \textit{zero-impact} threshold, we search for secondary user precoders causing minimum degradation to the sum rate of the active users through a steepest descent search over the Grassmann manifold. For this numerical optimization, we propose three initial solutions of varying degrees of complexity. 
Our initial work in \cite{BehrangICASSP2011} only deals with the case of enough antennas at the secondary users. This paper elaborates on the claims of \cite{BehrangICASSP2011}, provides new analysis for the case of enough antennas at the secondary users, considers the case of not enough antennas at the secondary users, and presents new simulations results.

\indent The remainder of the paper is organized as follows. In Section \ref{sec:SystemModel} we present the system model and review IA in the $K$-user constant MIMO channel. Admission of the arriving users is considered in Section \ref{sec:SecondaryUserAdmission}. Section \ref{sec:ZeroInterfereceToPrimaryUsers} presents precoder designs for the secondary users with large number of transmit antennas. In Section \ref{sec:SecondaryUsersInterfereingWithTheActiveUsers} we find secondary user precoders when the secondary users do not have enough transmit antennas. Numerical results are presented in Section \ref{sec:NumericalResults} followed by concluding remarks in Section \ref{sec:Conclusion}.

\indent \textit{Notation}: Upper and lowercase bold letters represent matrices and vectors. $\mathbf{A}^*$, $\mathbf{A}^T$ and $\mathbf{A}^{-1}$ are conjugate transpose, transpose and inverse of $\mathbf{A}$. $\mathbf{A}(n,m)$ is the element on the $n$th row and the $m$th column of $\mathbf{A}$. $\rm{tr}\big(\mathbf{A}\big)$ and ${\rm{rank}}(\mathbf{A})$ are trace and rank of $\mathbf{A}$. $|| \cdot ||_F$ and $|| \cdot ||_0$ are the Frobenius and zero norms. $[\mathbf{A},\mathbf{B}]$ is the matrix constructed by horizontal concatenation of matrices $\mathbf{A}$ and $\mathbf{B}$.  $\lceil a \rceil$ is the smallest integer larger than or equal to $a$. Kronecker product is represented with $\otimes$. For two matrices $\mathbf{A}$ and $\mathbf{B}$ with dimensions $a_1\times a_2$ and $b_1\times b_2$, $\mathbf{A}\oplus\mathbf{B}$ is a block-diagonal matrix of dimension $(a_1+b_1) \times (a_2+b_2)$ with matrices $\mathbf{A}$ and $\mathbf{B}$ on its main diagonal. $\mathbf{I}_N$ is the $N\times N$ identity matrix, $\mathbf{0}_{a\times b}$ is the matrix of dimension $a\times b$ with all elements equal to zero. Finally, $\mathcal{O}$ is the set of all matrices with orthonormal columns.

\section{System Model}\label{sec:SystemModel}
\begin{figure}[t]
 \centering
\includegraphics[width=3.5in]{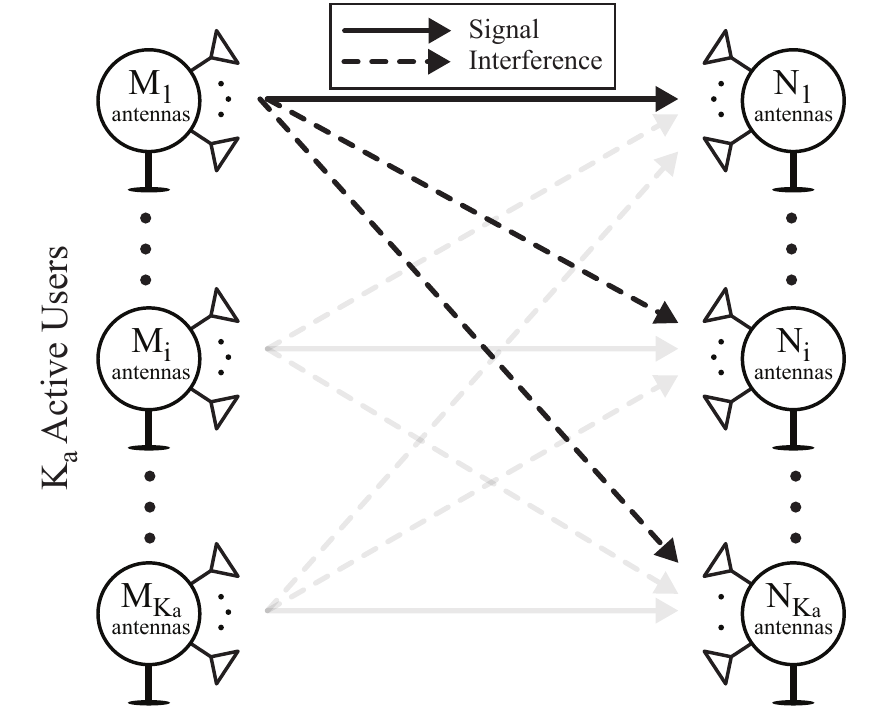}
\caption{$K_a$-user MIMO interference channel where the $i$th transmit/receive pair is equipped with $M_i$ transmit and $N_i$ receive antennas.}
\label{fig:SystemModel}
\end{figure}

\indent Consider a $K_a$-user MIMO interference channel shown in Fig. \ref{fig:SystemModel} where the $i$th transmitter and receiver pair are equipped with $M_i$ and $N_i$ antennas. Each transmitter $i$ uses a precoding matrix $\mathbf{F}_i$ of dimension $M_i\times d_i$ to transmit $d_i$ streams to its corresponding receiver. In this paper we assume $\mathbf{F}_i\in\mathcal{O}$. Note that precoders with orthonormal columns are useful for limited feedback codebook design \cite{love2004value}, in many cases unitary precoders are optimal when there exist a peak power constraint \cite{love2005}, and waterfilling over the eigenmodes of the effective channels can always be used to further improve the performance of the system after appropriate unitary precoders have been found. 
For each time instant, the received signal at the $i$th receiver, with perfect timing and synchronization, is
\begin{align}
\mathbf{y}_{i} = \sum_{k=1}^{K_a} \mathbf{H}_{ik}\mathbf{F}_{k}\mathbf{x}_{k} + \mathbf{z}_{i} 
\quad i=1,\ldots,K \, \label{eq:GeneralReceivedSignal}
\end{align}
where $\mathbf{H}_{ik}$ is the matrix of channel coefficients of a block fading channel between transmitter $k$ and receiver $i$, the transmitted signal from the $i$th node is $\mathbf{x}_i$ with power constraint $\mathbb{E}\{\mathbf{x}_i^*\mathbf{x}_i\} = P$ and $\mathbf{z}_i$ is the AWGN with elements in $\mathcal{CN}(0,\sigma^2)$ where $\sigma^2$ is the noise power spectral density which includes thermal noise and white excess interference from unaccounted sources. Note that similar to \cite{Peters2011}, we could incorporate colored noise by prewhitening or modifying the proposed algorithms but this is not considered in this pap er. 

\indent A $K_a$-user system of IA is feasible if there exists a set of matrices \(\mathcal{W}=\{\mathbf{W}_1,\ldots,\mathbf{W}_{K_a}\}\) such that, given the system model of \eqref{eq:GeneralReceivedSignal}, the following constraints are met \cite{Cadambe2008a}
\begin{align}
\left\{ \begin{array}{l}
{\rm{rank}}\left(\mathbf{W}_i\mathbf{H}_{ii}\mathbf{F}_i\right) = d_i \\
\mathbf{W}_i\mathbf{H}_{ik}\mathbf{F}_{k} = \mathbf{0}\ \forall k\neq i
\end{array} \right. \forall i,k\in\{1,\ldots,K_a\}, \label{eq:IAconds}
\end{align}
where $d_i$ is the number of interference-free streams the $i$th active transmitter can send to its receiver. The linear equalizer presented in \cite{BehrangISIT2010} and the projection matrix presented in \cite[Section III.A]{Peters2011} are examples of a possible receive filter $\mathbf{W}_i$ in \eqref{eq:IAconds}. We assume $\mathbf{H}_{ik}$ has full column rank for all $i,k$ and the set of $\{M_i, N_i,d_i,K_a\}$, $\forall i$, is feasible \cite{Yetis2010}. Through IA, interference from the active transmitters at the $i$th active receiver is confined to an $N_i-d_i$ dimensional subspace.
We denote the non-unique basis for the interference subspace as $\mathbf{C}_i$.
We assume the $i$th active transmitter sends $d_i$ streams and each active receiver uses a ZF equalizer, given by 
$\mathbf{W}_i\!=\!\left[\mathbf{I}_{d_i},\mathbf{0}_{d_i,N_i-d_i}\right]
[\mathbf{H}_{ii}\mathbf{F}_i,\mathbf{C}_i]^{-1}$ \cite{BehrangISIT2010}.
The total achievable sum rate of the active users is \cite{BehrangISIT2010}
\begin{align}
R_{{\rm{sum}}}^{{\rm{au}}} = \sum_{i=1}^{K_a} \sum_{n=1}^{d_i}\log\left(
1+\frac{\gamma_o/d_i}
{\mathbf{e}_n\left(\left(\mathbf{F}_i^*\mathbf{H}_{ii}^*\mathbf{P}_i
\mathbf{H}_{ii}\mathbf{F}_i\right)^{-1}\right)\mathbf{e}_n^*}\right), \label{eq:AchievableSumrateOfPrimaryUsersInAbsenceOfNU}
\end{align}
where $\mathbf{P}_i=(\mathbf{I}_{N_i}-\mathbf{C}_i\mathbf{C}_i^*)$ is the projection matrix into the nullspace of the interference subspace at the $i$th active receiver, $\gamma_o=\frac{P}{\sigma^2}$ and $\mathbf{e}_n$ is the $n$th row of $\mathbf{I}_n$. 
Note that the assumption of a ZF receiver at the active receivers is for mathematical tractability and developing intuition into the problem. The design of section \ref{subsec:IL_Min} is independent of the receiver type and the design of Section \ref{subsec:AM} can be extended to any receiver type that satisfies the rank constraint in \ref{eq:IAconds}. We assume global perfect CSI knowledge and that the secondary users can use the CSI of the active users' network to find the precoders/equalizers used by the active users. Although this assumption is not practical, the performance achieved by the  developed techniques in this paper can be used as a benchmark for the case that CSI is relayed via limited feedback \cite{Krishnamachari2009,Omar2010_AnalogConf}.

\section{Secondary User Admission}\label{sec:SecondaryUserAdmission}
Assume $K_s$ secondary users request access to the interference alignment group. Define $K=K_a+K_s$, and let the secondary users be the last $K_s$ users in the ordered set of user indices $\mathcal{K}=\{1,\ldots,K_a,K_a+1,\ldots,K\}$. The received signal at the $i$th $\leq K_a$ receiver is
$
\mathbf{y}_{i} = \sum_{k=1}^{K_a} \mathbf{H}_{ik}\mathbf{F}_{k}\mathbf{x}_{k}+
\sum_{\ell=K_a+1}^{K}\mathbf{H}_{i\ell}\mathbf{F}_{\ell}\mathbf{x}_{\ell}
+\mathbf{z}_{i}.$
We assume the first $K_a$ users do not change their precoders or receiving filters when the secondary users join the network. This has two important implications: \textit{$\left.a\right)$} the active transmitters do not help the transmission of the secondary users, \textit{$\left.b\right)$} the active receivers are not aware of the interference from the secondary users.  

\indent In the presence of the secondary users, the sum rate of the first $K_a$ users changes from \eqref{eq:AchievableSumrateOfPrimaryUsersInAbsenceOfNU} to
\begin{align}
R_{{\rm{sum}}}^{{\rm{au}}}\!=\!\sum_{i=1}^{K_a}\sum_{n=1}^{d_i}
\log\!\left(\! 1\! + \frac{\gamma_o/d_i}
{\mathbf{e}_n\left(\left(\mathbf{F}_i^*\mathbf{H}_{ii}^*\mathbf{P}_i
\mathbf{H}_{ii}\mathbf{F}_i\right)^{-1}\!
+\sum_{k=K_a+1}^{K}\frac{\gamma_o}{d_k}\mathbf{W}_i\mathbf{H}_{ ik}\mathbf{F}_{k}\mathbf{F}_{k}^*\mathbf{H}_{ik}^*\mathbf{W}_i^*\right)\mathbf{e}_n^* }
\right),
\label{eq:AchievableSumrateOfCUInPresenceOfNU}
\end{align}
where we have assumed that $\mathbf{H}_{ik}$ is independent of $\mathbf{H}_{nm}$ for $i\neq n$ or $k\neq m$. 
Under the assumptions that the active users are ignorant of the secondary users, a reasonable objective is for the secondary users to not degrade the sum rate of the first $K_a$ users. Comparing \eqref{eq:AchievableSumrateOfCUInPresenceOfNU} and \eqref{eq:AchievableSumrateOfPrimaryUsersInAbsenceOfNU} implies that $\sum_{k=K_a+1}^{K}\frac{\gamma_o}{d_k}\mathbf{W}_i\mathbf{H}_{ ik}\mathbf{F}_{k}\mathbf{F}_{k}^*\mathbf{H}_{ik}^*\mathbf{W}_i^*$ should ideally be zero or equivalently
\begin{align}
\mathbf{W}_i\mathbf{H}_{ik}\mathbf{F}_{k} = \mathbf{0}\ \ i\in\{1,\ldots,K_a\}, k\in\{K_a+1,\ldots,K\}. \label{eq:temp_ZeroInterFromCognitive}
\end{align}
It can be inferred from \eqref{eq:temp_ZeroInterFromCognitive} that the interference from the secondary users should be confined to the interference subspace of the active users \cite{BehrangISIT2010}. Therefore, by requiring that
$
\mathbf{P}_i \mathbf{H}_{ik}\mathbf{F}_{k} = \mathbf{0}
$ for $i\in\{1,\ldots,K_a\}$ and $k\in\{K_a+\!1,\ldots,K\}$, the precoder of the $k$th transmitter should satisfy
\begin{align}
\tilde{\mathbf{H}}_{k}\mathbf{F}_{k} = \mathbf{0}_{(\sum_{i=1}^{K_a} N_i)\times d_{k}}, \label{eq:NoInterferenceNullSpaceCond}
\end{align}
where $\tilde{\mathbf{H}}_{k}= \left[\left(\mathbf{P}_1\mathbf{H}_{1k}\right)^*,\ldots,\left(\mathbf{P}_{K_a}\mathbf{H}_{K_ak}\right)^*\right]^*$. Using \eqref{eq:NoInterferenceNullSpaceCond}, Lemma \ref{lem:MinimumNumberOfAntennasAtNUs} provides the required minimum number of antennas at each secondary user, called the \textit{zero-impact} threshold, which enables the secondary nodes to communicate, potentially, without affecting the sum rate of the active users.

\begin{Lem}\label{lem:MinimumNumberOfAntennasAtNUs}
The $k$th secondary user can transmit $d_k$ streams without degrading \eqref{eq:AchievableSumrateOfPrimaryUsersInAbsenceOfNU} if
\begin{align}
M_{k} \geq \sum_{i=1}^{K_a}d_i+d_{k}\quad k\in\{K_a+1,\ldots,K\}. \label{eq:NumberOfNUTransmitterAntennas_ZeroInterferenceToCU}
\end{align}
\end{Lem}
\begin{IEEEproof}
As $\mathbf{H}_{ik}$ has full column rank and $\mathbf{P}_i$ is a projection matrix into a $d_i$ dimensional subspace (${\rm{rank}}(\mathbf{P}_i) = d_i$), ${\rm{rank}}(\mathbf{P}_i\mathbf{H}_{ik}) = d_i$ \cite[Property 3.9.b]{Seber2008}. Hence, an $\mathbf{F}_k\neq\mathbf{0}$ satisfying \eqref{eq:NoInterferenceNullSpaceCond} can be found if $M_{k}>{\rm{rank}}(\tilde{\mathbf{H}}_{k}) = \sum\limits_{i=1}^{K_a}d_i$. As each secondary transmitter sends $d_{k}$ streams to its receiver, $\mathbf{F}_{k}$ (the nullspace of $\tilde{\mathbf{H}}_{k}$) should have rank of $d_{k}$ and the result follows.  
\end{IEEEproof}

\indent Based on Lemma \ref{lem:MinimumNumberOfAntennasAtNUs}, we divide the technical contributions of this paper into two parts.
\begin{enumerate}[i)]
\item The number of transmit antennas at each secondary user satisfies \eqref{eq:NumberOfNUTransmitterAntennas_ZeroInterferenceToCU}. In this case, the secondary users can also optimize some performance criterion for their own link.
\item The number of transmit antennas at each secondary user does not satisfy \eqref{eq:NumberOfNUTransmitterAntennas_ZeroInterferenceToCU}. The secondary users once admitted degrade the sum rate of the active users. 
\end{enumerate}
In the next two sections, we derive optimum and suboptimum secondary user precoders for each scenario based on $K_s$ and the available transmit antennas at the secondary users. From this point on, for mathematical tractability and to simplify the exposition, we assume the first $M_a\times N_a$ $K_a$ users each transmit $d_a$ streams and the last $M_s\times N_s$ $K_s$ users each transmit $d_s$ streams.   

\section{Secondary Users Not Interfering with The Active Users} \label{sec:ZeroInterfereceToPrimaryUsers}
\indent Based on \eqref{eq:NoInterferenceNullSpaceCond} and \eqref{eq:NumberOfNUTransmitterAntennas_ZeroInterferenceToCU}, the secondary transmitters need at least $K_ad_a+d_s$ antennas in order not to decrease the sum rate of the active receivers. This means each secondary user has to generate a precoder of rank $d_s$ using $M_s-K_ad_a$ bases. Therefore, when $M_s-K_ad_a>d_s$, the secondary users can use the extra degrees of freedom in designing their precoders to not only satisfy \eqref{eq:NoInterferenceNullSpaceCond} but also improve the performance of their own link. For the rest of this section, we assume $M_s > K_ad_a+d_s$ and that the secondary users want to maximize their sum rate given by
\begin{align}
R_{{\rm{sum}}}^{{\rm{su}}} =\negthickspace\negthickspace \sum_{k=K_a+1}^{K} \negthickspace\negthickspace\log\det\left(\mathbf{I}+\frac{\gamma_o}{ d_s}\mathbf{H}_{kk}
\mathbf{F}_k\mathbf{F}_k^*\mathbf{H}_{kk}^*\boldsymbol{\mathcal{I}}_k^{-1}\right), \label{eq:SumRateOfNUs_1}
\end{align}
where $\boldsymbol{\mathcal{I}}_k\! =\! (\mathbf{I}_{N_k}\! +\!\!\! \sum\limits_{i=1,i\neq k}^{K}\!\frac{\gamma_o}{d_i}\mathbf{H}_{k i}\mathbf{F}_{i}\mathbf{F}_{i}^*\mathbf{H}_{ki}^*)$ captures the interference from all the transmitters. 

\indent Let the right singular vectors of $\tilde{\mathbf{H}}_k$ spanning its nullspace be the columns of $\tilde{\mathbf{V}}_k$. It is inferred from \eqref{eq:NoInterferenceNullSpaceCond} that the columns of $\mathbf{F}_k$ have to be a linear combination of the columns of $\tilde{\mathbf{V}}_k$; i.e., $\mathbf{F}_k = \tilde{\mathbf{V}}_k\mathbf{G}_k$, for some matrix $\mathbf{G}_k$. As both $\mathbf{F}_k$ and $\tilde{\mathbf{V}}_k$ have orthonormal columns, columns of $\mathbf{G}_k$ are also orthonormal, i.e. $\mathbf{G}_k\in\mathcal{O}$. Using \eqref{eq:SumRateOfNUs_1}, the optimum secondary user precoders, $\hat{\mathbf{F}}_k=\tilde{\mathbf{V}}_k\hat{\mathbf{G}}_k$, are found by solving 
\begin{align}
\operatorname*{arg\,max}_{\mathbf{G}_{K_a+1},\ldots,\mathbf{G}_K\in \mathcal{O}} & \displaystyle \sum_{k=K_a+1}^{K} 
\log\det\left(\mathbf{I} + 
\frac{\gamma_o}{ d_s}\mathbf{H}_{kk}
\tilde{\mathbf{V}}_k\mathbf{G}_k\mathbf{G}_k^*\tilde{\mathbf{V}}_k^*
\mathbf{H}_{kk}^*\boldsymbol{\mathcal{I}}_k^{-1}\right),  \label{eq:OrigOptProblem_ObjectiveFunction_Plus_Constraint} 
\\
{\rm{s.t.}} \quad & \tilde{\mathbf{H}}_k\tilde{\mathbf{V}}_k\mathbf{G}_k = \mathbf{0}\quad k=K_a+1,\ldots,K.  \nonumber 
\end{align}
\noindent Solving \eqref{eq:OrigOptProblem_ObjectiveFunction_Plus_Constraint} is difficult in the most general case; we focus on several important special cases.

\subsection{Single New User} \label{subsec:SingleNewUserWithEnoughAntennas}
In this section we assume a single secondary user is arriving to the network. Then the optimum secondary user precoder solving \eqref{eq:OrigOptProblem_ObjectiveFunction_Plus_Constraint} can be found in closed-form and is given by Lemma \ref{Lem:ZeroInterferenceSumRateMaximizingCognitivePrecoding}.
\begin{Lem}\label{Lem:ZeroInterferenceSumRateMaximizingCognitivePrecoding} 
For $K_s=1$, the columns of $\hat{\mathbf{G}}_k$ solving \eqref{eq:OrigOptProblem_ObjectiveFunction_Plus_Constraint} 
are the $d_s$ most significant eigenvectors of $\tilde{\mathbf{V}}_k^*\mathbf{H}_{kk}^*\boldsymbol{\mathcal{I}}_k^{-1}
\mathbf{H}_{kk}\tilde{\mathbf{V}}_k$.
\end{Lem}
\begin{IEEEproof}
Let the summands in the objective function of \eqref{eq:OrigOptProblem_ObjectiveFunction_Plus_Constraint} be $f(\mathbf{G}_k)$. Using the identity $\det\left(\mathbf{I}\!+\!\mathbf{A}\mathbf{B}\right) = \det\left(\mathbf{I}\!+\!\mathbf{B}\mathbf{A}\right)$ for any two matrices $\mathbf{A}$ and $\mathbf{B}$, rewrite $f(\mathbf{G}_k)$ as 
$$f(\mathbf{G}_k) = \log\det \left(\mathbf{I}+\frac{\gamma_o}{d_s}\tilde{\mathbf{V}}_k^*
\mathbf{H}_{kk}^*\boldsymbol{\mathcal{I}}_k^{-1}\mathbf{H}_{kk}\tilde{\mathbf{V}}_k\mathbf{G}_k
\mathbf{G}_k^* \right).$$ 
Let the eigenvalue decomposition of $\tilde{\mathbf{V}}_k^*\mathbf{H}_{kk}^*\boldsymbol{\mathcal{I}}_k^{-1}\mathbf{H}_{kk}\tilde{\mathbf{V}}_k$ be $\mathbf{U}_k\mathbf{\Sigma}_k\mathbf{U}_k^*$ where $\mathbf{\Sigma}_k$ is a diagonal matrix holding its eigenvalues. Then 
$
f(\mathbf{G}_k) = \log\det \left(\mathbf{I}+\frac{\gamma_o}{d_s}\mathbf{\Sigma}_k\mathbf{U}_k^*\mathbf{G}_k
\mathbf{G}_k^*\mathbf{U}_k \right) 
$, and using the Hadamard inequality, the $\mathbf{G}_k$ diagonalizing the term inside the $\det$ maximizes $f(\mathbf{G}_k)$. A solution is found by setting the columns of $\mathbf{G}_k$ to the $d_s$ most significant eigenvectors in $\mathbf{U}_k$.
\end{IEEEproof}
If the secondary user was not optimizing its own link, $\mathbf{G}_k$ could be set to a random unitary matrix. Call this approach \textit{random orthonormal} design. Moreover, the secondary user could ignore the IA nodes and maximize its own achievable rate by adopting a \textit{selfish} precoding through setting the columns of its precoder to the $d_s$ most significant eigenvectors of 
$\mathbf{H}_{KK}^*\left(\mathbf{I}_{N_s} + \frac{\gamma_o}{d_a}\sum\limits_{i=1}^{K_a}\mathbf{H}_{Ki}\mathbf{F}_i\mathbf{F}_i^*\mathbf{H}_{Ki}^* \right)^{-1}\mathbf{H}_{KK}$.
Note that generalizing Lemma \ref{Lem:ZeroInterferenceSumRateMaximizingCognitivePrecoding} for $K_s>1$ is not trivial while random orthonormal and selfish precoding can be used for arbitrary $K_s$.

\subsection{Two New Users}\label{subsec:TwoNewUserWithEnoughAntennas}
Now assume $K_s=2$ where the goal of the secondary users is satisfying \eqref{eq:NoInterferenceNullSpaceCond} while maximizing the total achievable sum rate of the secondary users given by \eqref{eq:SumRateOfNUs_1}. In this case, $\boldsymbol{\mathcal{I}}_k$ in \eqref{eq:SumRateOfNUs_1} is
\begin{align}
\boldsymbol{\mathcal{I}}_k = \mathbf{I}+
\frac{\gamma_o}{d_a}\sum_{i=1}^{K_a}\mathbf{H}_{ki}\mathbf{F}_{i}\mathbf{F}_{i}^*
\mathbf{H}_{ki}^*+ \frac{\gamma_o}{d_s}\mathbf{H}_{kq}\mathbf{F}_{q}\mathbf{F}_{q}^*\mathbf{H}_{kq}^*
 \quad q, k\in\{K-1,K\}, \label{eq:BkInCaseOfTwoNU}
\end{align}
where $q\neq k$. Corollaries \ref{Col:ForgettingAboutTheSecondGuy} and \ref{Col:SuccessiveForTwoUsers} give approximate solution for \eqref{eq:OrigOptProblem_ObjectiveFunction_Plus_Constraint} in some special cases.

\begin{Col}\label{Col:ForgettingAboutTheSecondGuy}
When $K_a\gg (K_s-1) \geq 1$, \eqref{eq:OrigOptProblem_ObjectiveFunction_Plus_Constraint} is solved by setting the columns of $\mathbf{G}_{k}$ to the $d_s$ most significant eigenvectors of 
$
\tilde{\mathbf{V}}_k^*
\mathbf{H}_{kk}^*\left(\mathbf{I}_{N_s} + \frac{\gamma_o}{d_a}\sum_{i=1}^{K_a}\mathbf{H}_{k i}\mathbf{F}_{i}\mathbf{F}_{i}^*\mathbf{H}_{ki}^*\right)^{-1}\mathbf{H}_{kk}
\tilde{\mathbf{V}}_k$.
\end{Col}
\begin{IEEEproof}
When $K_a\gg (K_s-1) \geq 1$, the second term in \eqref{eq:BkInCaseOfTwoNU}, representing the interference from the active transmitters, dominates the third term which is the interference from the other secondary user. Ignoring the third term in \eqref{eq:BkInCaseOfTwoNU} changes the optimization of \eqref{eq:OrigOptProblem_ObjectiveFunction_Plus_Constraint} to the case of $K_s=1$ which is solved using Lemma \ref{Lem:ZeroInterferenceSumRateMaximizingCognitivePrecoding} and the result follows. 
\end{IEEEproof}
\indent In Corollary \ref{Col:ForgettingAboutTheSecondGuy}, the secondary users can independently design their precoders. A higher sum rate is obtainable, however, by successive application of Lemma \ref{Lem:ZeroInterferenceSumRateMaximizingCognitivePrecoding} to the nodes in the ordered list of $\{K_a+1,\ldots,K\}$ such that the $k$th secondary user ignores the interference from the $i$th transmitter for $i\in\{k+1,\ldots,K\}$. Corollary \ref{Col:SuccessiveForTwoUsers} gives a more precise description of this approach where for conciseness we ignore the gain attainable through ordering the users.

\begin{Col}\label{Col:SuccessiveForTwoUsers}
When $K_a\gg (K_s-1) \geq 1$, \eqref{eq:OrigOptProblem_ObjectiveFunction_Plus_Constraint} is approximately solved by computing $\mathbf{G}_{K-1}$ using Corollary \ref{Col:ForgettingAboutTheSecondGuy} and setting the columns of $\mathbf{G}_{K}$ to the $d_s$ most significant eigenvectors of 
\begin{align}
&\tilde{\mathbf{V}}_{K}^*
\mathbf{H}_{KK}^*\left(\mathbf{I}_{N_s} + \frac{\gamma_o}{d_s}\mathbf{H}_{K(K-1)}\hat{\mathbf{F}}_{K-1}\hat{\mathbf{F}}_{K-1}^*\mathbf{H}_{K(K-1)}^* + 
\frac{\gamma_o}{d_a}\sum_{i=1}^{K_a}\mathbf{H}_{K i}\mathbf{F}_{i}\mathbf{F}_{i}^*\mathbf{H}_{Ki}^*\right)^{-1}\mathbf{H}_{KK}
\tilde{\mathbf{V}}_{K}.
\nonumber 
\end{align}
\end{Col}

\indent In an IA network configured for $K_a$ users, the number of secondary users arriving is expected to be less than the number of users currently in the network which justifies the assumptions of Corollaries \ref{Col:ForgettingAboutTheSecondGuy} and \ref{Col:SuccessiveForTwoUsers}. Moreover, both Corollaries \ref{Col:ForgettingAboutTheSecondGuy} and \ref{Col:SuccessiveForTwoUsers} can be readily extended to more than $K_s=2$. As $K_s$ increases, the interference from the other secondary user(s) becomes non-negligible and we expect the performance difference between Corollaries \ref{Col:ForgettingAboutTheSecondGuy} and \ref{Col:SuccessiveForTwoUsers} to increase.

\indent In Corollary \ref{Col:SuccessiveForTwoUsers}, one of the secondary users interferes with the other one. A more involved successive application of Lemma \ref{Lem:ZeroInterferenceSumRateMaximizingCognitivePrecoding} can confine the interference from the $K$th transmitter at the $(K-1)$th receiver to the $(N_{K-1}- \sum_{i=1}^{K_a}d_i)$ dimensional subspace spanned by the interference from the active users which implies $M_K \geq \sum\limits_{i=1}^{K_a}d_i + (N_{K-1}- \sum\limits_{i=1}^{K_a}d_i)+ d_K  = N_{K-1} + d_K$.

\subsection{More Than Two New Users} \label{subsec:MoreThanTwoNUs}
Solving \eqref{eq:OrigOptProblem_ObjectiveFunction_Plus_Constraint} for $K_s\geq2$ is equivalent to solving the general sum rate maximizing precoder of the MIMO interference channel and, to date, a closed-form solution directly solving it does not exist \cite{Matskani2008}. As an alternative, we provide a precoder design maximizing the pre-log factor of \eqref{eq:SumRateOfNUs_1} at asymptotically high transmit power for a certain (and plausible) network configuration.

\indent From \eqref{eq:NoInterferenceNullSpaceCond}, the $i$th secondary transmitter has to allocate $K_ad_a$ of its spatial dimensions to pre-align the interference it causes to the active receivers by selecting its precoder
in the column space of $\tilde{\mathbf{V}}_{i}$. Interference from the active transmitters at the $k$th secondary receiver spans a $K_ad_a$ dimensional subspace given by the \textit{interference basis} of $\{\mathbf{H}_{k1}\mathbf{F}_1,\ldots,\mathbf{H}_{kK_a}\mathbf{F}_{Ka}\}$. Let columns of $\tilde{\mathbf{W}}_{k}$ span the left nullspace of \begin{math}\left[(\mathbf{H}_{k 1}\mathbf{F}_{1})^T,\ldots, (\mathbf{H}_{k K_a}\mathbf{F}_{K_a})^T\right]^T\end{math}. An equalizer at the $k$th secondary receiver post multiplied by $\tilde{\mathbf{W}}_{k}$ always cancels the interference from the active transmitters. Let the channels between the secondary users, $\mathbf{H}_{ki}$ for $k,i\in\{K_a+1,\ldots,K\}$, be replaced by $(M_s-K_ad_a)\times(N_s-K_ad_a)$ dimensional effective channels of $\tilde{\mathbf{W}}_{k}\mathbf{H}_{ki}\tilde{\mathbf{V}}_{i}$. Then the networks of the secondary and active users will become disjointed meaning that the secondary nodes will not cause/receive any interference to/from the active nodes. 

\indent Assume $M_s = N_s$ and let $\tilde{M_s}=\tilde{N_s}=M_s-K_ad_a$. If $K_s \leq 2\frac{\tilde{M_s}}{d_s} -1$ \cite{Yetis2010}, or $\tilde{M}_s = \lceil\frac{1+K_s}{2}\rceil$ (when $d_s=1$), we can perform another level of IA between the secondary users over their effective channels. Hence both the active and the secondary users will be performing IA among themselves and while the interference from the secondary transmitters is pre-aligned to the interference subspaces at the active receivers, the interference from the active transmitters is always canceled at the secondary receivers. We call this method \textit{successive IA} and conjecture that in special cases, it achieves the same DOF as if all the $K$ users 
had performed IA together.

\begin{Conj}\label{Conj:MultiLeveLIA}
Consider a $3K$-user interference channel for $K\in \mathbb{Z}^+$ where the transmitter/receiver pairs are divided into $K$ groups of $3$ users each, $\left\{G_1,G_2,\ldots,G_K\right\}$, such that the nodes of the $i$th group have $3i-1$ antennas. Performing successive IA on the $k$th group, $2 \leq k\leq K$, through creating effective channels between the nodes of $G_k$ based on the interference subspaces and precoders of the $\{1,\ldots,k-1\}$ groups achieves the same DOF as if all the $3K$ users had done IA together.
\end{Conj}

\indent Using counting arguments, it is possible to show that successive IA achieves $3K$ DOF. Proving that IA can only achieve $3K$ DOF, however, is not trivial. We have numerically confirmed (up to $K=5$) that only $3K$ DOF is achievable if all the nodes perform IA together (for $\hbox{DOF} >K$, the algorithm of \cite[Section III.A]{Peters2011} does not converge). We also use \cite[Theorem 2]{Yetis2010} to further support Conjecture \ref{Conj:MultiLeveLIA}. Let all the nodes except the $\ell$th node (in the $i$th group) transmit a single stream. Then, the number of variables ($N_v$) and the number of equations ($N_e$) in 
\eqref{eq:IAconds} are 
\begin{align}
N_v &= \sum_{k=1}^{3K}d_k(M_s+N_k-2d_k) = 9K^2 -3K -6i+4 + d_{\ell}(6i-2-2d_{\ell}), \label{eq:N_v}
\\
N_e &= \sum_{k=1}^{3K-1}\left(d_{\ell} + 3K-2\right) + d_{\ell}\left(3K-1\right) = \left(3K-1\right)\left(3K-2+2d_{\ell}\right). \label{eq:N_e}
\end{align}
From \eqref{eq:N_v} and \eqref{eq:N_e}, since $d_{\ell}>1$ and $i\leq K$, this system is always improper ($N_v<N_e$)
\begin{align}
N_v  = 9K^2 -3K -6i+4 + d_{\ell}(6i-2-2d_{\ell}) &< N_e = \left(3K-1\right)\left(3K-2+2d_{\ell}\right) \nonumber \\
6d_{\ell}(i-K) - 2d_{\ell}^2 -6(i+K) &< -2. \nonumber 
\end{align}

\indent Note that if $K_s>2\frac{\tilde{M}_s}{d_s} -1$ or $\tilde{M}_s\neq\lceil\frac{1+K_s}{2}\rceil$ (assuming $M_s=N_s$), 
successive IA is not possible and, in general, the secondary users precoders have to be found through numerical optimization methods. Secondary users, however, can use suboptimal precoding designs including:
\begin{itemize}
\item Each secondary user independently designs its precoder based on Corollary \ref{Col:ForgettingAboutTheSecondGuy}
\item Secondary users successively design their precoders based on Corollary \ref{Col:SuccessiveForTwoUsers}. 
\item Performing successive IA on a subset of secondary users and using any combination of the previous suboptimum methods for the rest of the secondary transmitters. 
\end{itemize}
We present numerical comparison of these methods in Section \ref{sec:NumericalResults} and leave analytical comparison of such suboptimum solutions for future work.

\section{Secondary Users Interfering with the Active Users} \label{sec:SecondaryUsersInterfereingWithTheActiveUsers}
In this section we assume that the number of transmit antennas at the secondary users is less than the \textit{zero-impact} threshold given by \eqref{eq:NumberOfNUTransmitterAntennas_ZeroInterferenceToCU}. This implies that  \eqref{eq:temp_ZeroInterFromCognitive} cannot be satisfied and the secondary users are bound to change the sum rate of the active users. Consequently, we desire the secondary users to cause minimum degradation to the sum rate of the active users given by \eqref{eq:AchievableSumrateOfCUInPresenceOfNU}. Defining $\mathbf{Q}_i=(\mathbf{F}_i^*\mathbf{H}_{ii}^*\mathbf{P}_i
\mathbf{H}_{ii}\mathbf{F}_i)^{-1}$, we seek a solution to the problem of
\begin{eqnarray}
\operatorname*{arg\,max}_{\mathbf{F}_{K_a+1},\ldots,\mathbf{F}_K\in\mathcal{O}} &\displaystyle \sum_{i=1}^{K_a}\sum_{n=1}^{d_i}
\log\left(\!\!1\! +\! \frac{\gamma_o/d_i}
{\mathbf{e}_n\left(\mathbf{Q}_i
+\frac{\gamma_o}{d_s}\mathbf{W}_i\left(\sum_{k=K_a+1}^{K}\mathbf{H}_{ ik}\mathbf{F}_{k}\mathbf{F}_{k}^*\mathbf{H}_{ik}^*\right)\mathbf{W}_i^*\right)\mathbf{e}_n^* }
\right). \label{eq:SecondOptProblem}
\end{eqnarray}

\indent Like the sum rate maximizing interference alignment solution, solving \eqref{eq:SecondOptProblem} in its most general form is challenging. Consequently we pursue a numerical solution for certain special cases. Assuming $K_s=1$, \eqref{eq:SecondOptProblem} simplifies to
\begin{align}
\operatorname*{arg\,max}_{\mathbf{F}_s\in\mathcal{O}}\sum_{i=1}^{K_a}\sum_{n=1}^{d_i}
\log\left( 1 + \frac{\gamma_o/d_i}
{\mathbf{e}_n\left(\mathbf{Q}_i
+\frac{\gamma_o}{d_s}\mathbf{W}_i\mathbf{H}_{ is}\mathbf{F}_{s}\mathbf{F}_{s}^*\mathbf{H}_{is}^*\mathbf{W}_i^*\right)\mathbf{e}_n^* }
\right), \label{eq:OptimizingSumrateAtCUWithInterference_temp1}
\end{align}
where subscript $s$ refers to the single secondary user. As $\mathbf{F}_s\in\mathcal{O}$ and $R_{\rm{sum}}^{\rm{au}}(\mathbf{F}_s) = R_{\rm{sum}}^{\rm{au}}(\mathbf{F}_s\mathbf{Q})$ for any unitary matrix $\mathbf{Q}$, $R_{\rm{sum}}^{\rm{au}}$ has the same value for all the points in the range of $\mathbf{F}_s$. Equivalently, we are searching for a $d_s$-dimensional subspace of the vector space $\mathbb{C}^{M_s}$ and therefore, our search is confined to the Grassmann manifold. Hence, we can find locally optimum solutions for \eqref{eq:OptimizingSumrateAtCUWithInterference_temp1} using numerical optimization over the Grassmann manifold. We use the \textit{modified steepest descent in the complex Grassmann manifold} algorithm (MGM) in \cite{Manton2002} which requires the derivative of \eqref{eq:OptimizingSumrateAtCUWithInterference_temp1} with respect to $\mathbf{F}_s$ and an initial guess. The problem in \eqref{eq:OptimizingSumrateAtCUWithInterference_temp1} is not convex and MGM only guarantees convergence to a local optimum point. Therefore, a better initial solution will lead to a better final output of the MGM algorithm. Next we propose three initial solutions of varying degrees of complexity for the MGM algorithm.

\subsection{Alternating Minimization Based Suboptimum Solution} \label{subsec:AM}
Let $\tilde{\mathbf{W}}=\bigoplus\limits_{i=1}^{K_a}\bigoplus\limits_{n=1}^{d_i}\sqrt{d_i}\mathbf{e}_n\mathbf{W}_i\mathbf{H}_{is}$, $\mathbf{S}=\bigoplus\limits_{i=1}^{K_a}\bigoplus\limits_{n=1}^{d_i}d_i\mathbf{e}_n\mathbf{Q}_i\mathbf{e}_n^*$, and $\tilde{d} = K_ad_a$. At high SINR, we can neglect the $1$ and rewrite \eqref{eq:OptimizingSumrateAtCUWithInterference_temp1} as 
\begin{align}
&\operatorname*{arg\,max}_{\mathbf{F}_s\in\mathcal{O}}\ -\log\det\left(
\frac{1}{\gamma_o}\mathbf{S}+ \frac{1}{d_s}\left(\tilde{\mathbf{W}}(\mathbf{I}_{\tilde{d}}
\otimes\mathbf{F}_s\mathbf{F}_s^*)\tilde{\mathbf{W}}^*\right)\right), \label{eq:SumOfActiveUsersInPresenceOfaSingleSecondaryUserWithNotEnoughAntennas}
\\
= &\operatorname*{arg\,min}_{\mathbf{F}_s\in\mathcal{O}}\  \log\det\left(\mathbf{I} +
\frac{\gamma_o}{d_s}\mathbf{S}^{-1}\tilde{\mathbf{W}}(\mathbf{I}_{\tilde{d}}\otimes\mathbf{F}_s\mathbf{F}_s^*)\tilde{\mathbf{W}}^*\right).
\label{eq:NewSecondOptimizationProblem_temp}
\end{align}
where non-singularity of $\mathbf{S}$ (due to the rank constraint in \eqref{eq:IAconds}) was used to derive \eqref{eq:NewSecondOptimizationProblem_temp} from \eqref{eq:SumOfActiveUsersInPresenceOfaSingleSecondaryUserWithNotEnoughAntennas}. As $\mathbf{I}_{\tilde{d}}\otimes\mathbf{F}_s\mathbf{F}_s^* = \left(\mathbf{I}_{\tilde{d}}\otimes\mathbf{F}_s\right)\left(\mathbf{I}_{\tilde{d}}\otimes\mathbf{F}_s^*\right)$ and $\det\left(\mathbf{I}+\mathbf{A}\mathbf{B}\right) = \det\left(\mathbf{I}+\mathbf{B}\mathbf{A}\right)$ for any two matrices $\mathbf{A}$ and $\mathbf{B}$, \eqref{eq:NewSecondOptimizationProblem_temp} is equivalent to
\begin{align}
\operatorname*{arg\,min}_{\mathbf{F}_s\in\mathcal{O}}\
\log\det\left(
\mathbf{I} +
\frac{\gamma_o}{d_s}(\mathbf{I}_{\tilde{d}}\otimes\mathbf{F}_s^*)\tilde{\mathbf{W}}^{*}\mathbf{S}^{-1}\tilde{\mathbf{W}}(\mathbf{I}_
{\tilde{d}}\otimes\mathbf{F}_s)\right). \label{eq:NewSecondOptimizationProblem}
\end{align}
Let the eigenvalue decomposition of $\tilde{\mathbf{W}}^{*}\mathbf{S}^{-1}\tilde{\mathbf{W}}$ be $\mathbf{U}_{\tilde{\mathbf{W}}}\boldsymbol{\Sigma}_{\tilde{\mathbf{W}}}\mathbf{U}_{\tilde{\mathbf{W}}}^*$. Note that ${\rm{rank}}\left(\tilde{\mathbf{W}}^{*}\mathbf{S}^{-1}\tilde{\mathbf{W}}\right) = 
K_ad_a$. If the rows of $\left(\mathbf{I}_{\tilde{d}}\otimes\mathbf{F}_s^*\right)$ were equal to the linear combination of $(M_s-1)K_ad_a$ eigenvectors corresponding to zero eigenvalues 
in $\boldsymbol{\Sigma}_{\tilde{\mathbf{W}}}$, the $\det$ in \eqref{eq:NewSecondOptimizationProblem} would attain its minimum value of $1$. Therefore, ideally we would like to have
\begin{align}
\mathbf{I}_{\tilde{d}}\otimes\mathbf{F}_s = \tilde{\mathbf{U}}_{\tilde{\mathbf{W}}}\mathbf{A}_s, \label{eq:OriginalEquationForAs}
\end{align}  
where the columns of $\tilde{\mathbf{U}}_{\tilde{\mathbf{W}}}$ are the columns of $\mathbf{U}_{\tilde{\mathbf{W}}}$ corresponding to zero eigenvalues of $\tilde{\mathbf{W}}^{*}\mathbf{S}^{-1}\tilde{\mathbf{W}}$ and $\mathbf{A}_s$ is an $(M_s-1)\sum_{i=1}^{K_a}d_i\times (\sum_{i=1}^{K_a}d_i)d_s$ combining matrix such that
\begin{align}
\mathbf{A}_s^*\mathbf{A}_s = \mathbf{A}_s^*\tilde{\mathbf{U}}_{\tilde{\mathbf{W}}}^*\tilde{\mathbf{U}}_{\tilde{\mathbf{W}}}\mathbf{A}_s = (\mathbf{I}_{\tilde{d}}\otimes\mathbf{F}_s^*)(\mathbf{I}_{\tilde{d}}\otimes\mathbf{F}_s) = \mathbf{I}_{\tilde{d}}\otimes \mathbf{F}_s^*\mathbf{F}_s = \mathbf{I}_{\tilde{d}}\otimes \mathbf{I}_{d_s}
= \mathbf{I}_{d_s\tilde{d}}. \label{eq:UnitaryConditionOnAs}
\end{align} 

\indent Let $\tilde{\mathbf{U}}_{\tilde{\mathbf{W}}} = \left[\tilde{\boldsymbol{\mathcal{U}}}_1^T,\ldots,\tilde{\boldsymbol{\mathcal{U}}}_{\tilde{d}}^T\right]^T$, where  $\tilde{\boldsymbol{\mathcal{U}}}_{q}$ for $q=1,\ldots,\tilde{d}$ are row-blocks of dimension $M_s\times (M_s-1)\tilde{d}$. Let $\mathbf{A}_s = \left[\boldsymbol{\mathcal{A}}_1,\ldots,\boldsymbol{\mathcal{A}}_{\tilde{d}}\right]$, where $\boldsymbol{\mathcal{A}}_p$ for $p = 1,\ldots,\tilde{d}$ are column-blocks of dimension $(M_s-1)\tilde{d}\times d_s$. Considering the block diagonal structure of $\mathbf{I}_{\tilde{d}}\otimes \mathbf{F}_s$, \eqref{eq:OriginalEquationForAs} can be rewritten as
\begin{align}
\left\{\begin{array}{l}
\tilde{\boldsymbol{\mathcal{U}}}_p\boldsymbol{\mathcal{A}}_q = \mathbf{0}\quad p,q=1,\ldots,\tilde{d}, \ p\neq q\\
\tilde{\boldsymbol{\mathcal{U}}}_p\boldsymbol{\mathcal{A}}_p = \tilde{\boldsymbol{\mathcal{U}}}_{p+1}\boldsymbol{\mathcal{A}}_{p+1} \quad p=1,\ldots,\tilde{d}-1
\end{array}\right. . \label{eq:systemOfEquationsForAsMidWay_1}
\end{align}
The system of equations in \eqref{eq:systemOfEquationsForAsMidWay_1} can be further simplified to
$
\tilde{\boldsymbol{\mathcal{U}}}\boldsymbol{\mathbf{a}}_s = \mathbf{0}
$, 
where $\mathbf{a}_s = {\rm{vec}}\left(\mathbf{A}_s\right)$ and $\tilde{\boldsymbol{\mathcal{U}}}$ is found by vertically stacking the coefficients in the $(M_sd_s)\tilde{d}^2 - (M_sd_s)\tilde{d} + (M_sd_s)(\tilde{d} -1) = (M_sd_s)\left(\tilde{d}^2-1\right)$ equations given by \eqref{eq:systemOfEquationsForAsMidWay_1}. Note that as $\tilde{d}^2 \geq M_s$, the nullspace of $\tilde{\boldsymbol{\mathcal{U}}}$ only consists of the null vector and a non-zero $\mathbf{a}_s$ exactly solving \eqref{eq:systemOfEquationsForAsMidWay_1} cannot be found. Next we develop an alternating minimization algorithm to solve  \eqref{eq:OriginalEquationForAs} in the least squares sense. 

\indent Assume $\mathbf{F}_s$ in \eqref{eq:OriginalEquationForAs} is given. We seek an $\mathbf{A}_s$ to minimize
$||\mathbf{I}_{\tilde{d}}\otimes \mathbf{F}_s - \tilde{\mathbf{U}}_{\tilde{\mathbf{W}}}\mathbf{A}_s ||_F ^2$. Formally,
\begin{align}
\hat{\mathbf{A}}_s =\  &\operatorname*{arg\,min}_{\mathbf{A}_s\in\mathcal{O}} \
{\rm{tr}}\left(\left(\mathbf{I}_{\tilde{d}}\otimes \mathbf{F}_s - \tilde{\mathbf{U}}_{\tilde{\mathbf{W}}}\mathbf{A}_s\right)^*\left(\mathbf{I}_{\tilde{d}}\otimes \mathbf{F}_s - \tilde{\mathbf{U}}_{\tilde{\mathbf{W}}}\mathbf{A}_s\right)\right) \label{eq:GivenFsFindingAs_1} 
\\
 = \ &\operatorname*{arg\,min}_{\mathbf{A}_s\in\mathcal{O}} \ {\rm{tr}}\left( \mathbf{I}_{\tilde{d}}\otimes \mathbf{F}_s^*\mathbf{F}_s 
\!+\! \mathbf{A}_s^*\tilde{\mathbf{U}}_{\tilde{\mathbf{W}}}^*\tilde{\mathbf{U}}_{\tilde{\mathbf{W}}}\mathbf{A}_s
- (\mathbf{I}_{\tilde{d}}\otimes \mathbf{F}_s^*)\tilde{\mathbf{U}}_{\tilde{\mathbf{W}}}\mathbf{A}_s
-\mathbf{A}_s^*\tilde{\mathbf{U}}_{\tilde{\mathbf{W}}}^*(\mathbf{I}_{\tilde{d}}\otimes \mathbf{F}_s)
\right). \label{eq:GivenFsFindingAs_2} 
\end{align}
Using \eqref{eq:UnitaryConditionOnAs}, \eqref{eq:GivenFsFindingAs_2} simplifies to
\begin{align}
\hat{\mathbf{A}}_s\! = &\operatorname*{arg\,min}_{\mathbf{A}_s\in\mathcal{O}} \ {\rm{tr}}\left( \mathbf{I}_{d_s\tilde{d}} - \! \mathfrak{Re}\left(\! (\mathbf{I}_{\tilde{d}}\otimes \mathbf{F}_s^*)\tilde{\mathbf{U}}_{\tilde{\mathbf{W}}}\mathbf{A}_s\!\right)\!\right)
= \operatorname*{arg\,max}_{\mathbf{A}_s\in\mathcal{O}} \ \mathfrak{Re}\left(\!{\rm{tr}} \left(\!\mathbf{A}_s^*\tilde{\mathbf{U}}_{\tilde{\mathbf{W}}}^* (\mathbf{I}_{\tilde{d}}\otimes \mathbf{F}_s)\right)
\!\right), \label{eq:GivenFsFindingAs_3}
\end{align}
where $\mathfrak{Re}\{.\}$ selects the real part of a complex number. The optimization problem of \eqref{eq:GivenFsFindingAs_3} can be solved using the solution to the ``Procrustes problem" \cite{Schonemann1966}. Writing the singular value decomposition of $\tilde{\mathbf{U}}_{\tilde{\mathbf{W}}}^* (\mathbf{I}_{\tilde{d}}\otimes \mathbf{F}_s)$ as 
$\begin{bmatrix} \boldsymbol{\Phi}_1 & \boldsymbol{\Phi_2} \end{bmatrix}
\begin{bmatrix} \boldsymbol{\Sigma}_1^T & \mathbf{0}^T\end{bmatrix}^T\mathbf{\Delta}^*$, the solution to \eqref{eq:GivenFsFindingAs_3} is given by $\hat{\mathbf{A}}_s=\boldsymbol{\Phi}_1\mathbf{\Delta}^*$ with the maximum value of \eqref{eq:GivenFsFindingAs_3} equal to $\mathfrak{Re}\left({\rm{tr}}\left(\boldsymbol{\Sigma_1}\right)\right)$.

\indent Again consider \eqref{eq:OriginalEquationForAs} but assume that this time $\mathbf{A}_s$ is given. We seek 
\begin{align}
\hat{\mathbf{F}}_s &=\ \operatorname*{arg\,min}_{\mathbf{F}_s\in\mathcal{O}}  ||\mathbf{I}_{\tilde{d}}\otimes \mathbf{F}_s - \tilde{\mathbf{U}}_{\tilde{\mathbf{W}}}\mathbf{A}_s ||_F ^2 
\ = \ \operatorname*{arg\,max}_{\mathbf{F}_s\in\mathcal{O}} \ \mathfrak{Re}\left({\rm{tr}} \left((\mathbf{I}_{\tilde{d}}\otimes \mathbf{F}_s)\mathbf{A}_s^*\tilde{\mathbf{U}}_{\tilde{\mathbf{W}}}^* \right)
\right). \label{eq:GivenAsFindingFs_1_5}
\end{align}
Using the sparsity of $\mathbf{I}_{\tilde{d}}\otimes \mathbf{F}_s$, \eqref{eq:GivenAsFindingFs_1_5} can be written as
\begin{align}
\hat{\mathbf{F}}_s =\ \operatorname*{arg\,max}_{\mathbf{F}_s\in\mathcal{O}} \ \hbox{$\mathfrak{Re}\left({\rm{tr}} \left( \mathbf{F}_s\sum_{k=1}^
{\tilde{d}}\boldsymbol{\mathcal{A}}_k^*\tilde{\boldsymbol{\mathcal{U}}}_k^*\right)\right)$}, \label{eq:GivenAsFindingFs_2}
\end{align}
where $\boldsymbol{\mathcal{A}}_k$ and $\tilde{\boldsymbol{\mathcal{U}}}_k$ are defined in \eqref{eq:systemOfEquationsForAsMidWay_1}. Let $\boldsymbol{\Psi}\begin{bmatrix} \boldsymbol{\Pi}_1 & \mathbf{0} \end{bmatrix}
\begin{bmatrix}(\boldsymbol{\Lambda}_1^*)^T (\boldsymbol{\Lambda}_2^*)^T \end{bmatrix}^T$ be the singular value decomposition of $\sum_{k=1}^{\tilde{d}}\boldsymbol{\mathcal{A}}_k^*\tilde{\boldsymbol{\mathcal{U}}}_k^*$. The solution to \eqref{eq:GivenAsFindingFs_2} is $\hat{\mathbf{F}}_s = \boldsymbol{\Lambda}_1\boldsymbol{\Psi}^*$ where the objective function attains its maximum value of $\mathfrak{Re}\left({\rm{tr}}\left(\Pi_1\right)\right)$. Based on \eqref{eq:GivenFsFindingAs_3} and \eqref{eq:GivenAsFindingFs_2}, the alternating minimization based approximate solution to \eqref{eq:OriginalEquationForAs} is given by Algorithm \ref{Alg:AMforDetMin}.
\begin{algorithm}
\caption{Alternating minimization based suboptimum solution}
\label{Alg:AMforDetMin}
\begin{algorithmic}
\State Initialize $\mathbf{F}_s$
\Repeat
\State Find $\hat{\mathbf{A}}_s$ using \eqref{eq:GivenFsFindingAs_3}
\State Find $\hat{\mathbf{F}}_s$ using \eqref{eq:GivenAsFindingFs_2}
\Until{Convergence}
\end{algorithmic}
\end{algorithm}

\indent After Algorithm \ref{Alg:AMforDetMin} converges, $\tilde{\mathbf{U}}_{\tilde{\mathbf{W}}}\hat{\mathbf{A}}_s \backsimeq \bigoplus\limits_{\ell=1}^{\tilde{d}}\mathbf{F}_{\ell,s}$ and an initial solution for the MGM algorithm, approximately solving  \eqref{eq:OptimizingSumrateAtCUWithInterference_temp1}, can be found by restricting $\mathbf{F}_s$ to the set of precoders in $\{\mathbf{F}_{1,s},\ldots,\mathbf{F}_{\tilde{d},1}\}$. As the non-negative objective function of \eqref{eq:GivenFsFindingAs_1} reduces at each step of Algorithm \ref{Alg:AMforDetMin}, Algorithm \ref{Alg:AMforDetMin} always converges. Convergence to a globally optimum point, however, is not guaranteed as $\mathcal{O}$ is not a convex set. Detailed results on convergence of alternating minimization algorithms similar to Algorithm \ref{Alg:AMforDetMin} can be found in \cite{Tropp2005}.

\subsection{Interference Leakage Minimizing Suboptimum Solution} \label{subsec:IL_Min}
The alternating minimization based initial solution suffers from high computational complexity. A less complex (and less accurate) initial solution to the MGM algorithm can be found by revisiting the constraint on the secondary user precoders outlined in \eqref{eq:NoInterferenceNullSpaceCond}. Instead of requiring the secondary transmitters' interference to be confined in the interference subspace of each active receiver, we minimize (in the least squares sense) the \textit{interference leakage} caused by the secondary users through solving
$\displaystyle \operatorname*{arg\,min}_{\mathbf{F}_s\in\mathcal{O}} || \tilde{\mathbf{H}}_{s}\mathbf{F}_s||_F^2$. A solution to this problem is given by setting the columns of $\mathbf{F}_s$ to the $d_s$ least significant right singular vectors of $\tilde{\mathbf{H}}_{s}$.

\subsection{DOF-Preserving Suboptimum Solution}\label{subsec:DOF_Preserv}
In both the initial solutions of Section \ref{subsec:AM} and Section \ref{subsec:IL_Min}, interference from the secondary transmitter is not confined to the interference subspaces of the active receivers. Thus, the DOF of the active users' network is zero. By aligning the secondary transmitter's interference at some of the active receivers, however, we can achieve a non-zero DOF and hence a better performance at high $\gamma_o$. Specifically, as $\mathbf{F}_s$ can not satisfy \eqref{eq:NoInterferenceNullSpaceCond}, we minimize the number of interference subspace dimensions at the active receivers to which the interference from the secondary transmitter is not aligned which translates to solving $\displaystyle \operatorname*{arg\,min}_{\mathbf{F}_s\in\mathcal{O}} || \tilde{\mathbf{H}}_{s}\mathbf{F}_s||_0$. In general, this is a combinatorial problem \cite{candes2005decoding} and exact solutions are out of the scope of this manuscript. Instead, we search for the largest subset of active receivers, $\tilde{\cal{K}}_a$, such that the system of equations defined by $\mathbf{P}_i\mathbf{H}_{is}\mathbf{F}_s=\mathbf{0}$ for all $i\in \tilde{\cal{K}}_a$ has an exact solution (equivalently $M_s\geq \sum\limits_{i\in \tilde{\cal{K}}_a}d_i + d_s$). In this fashion, $\mathbf{F}_s$ will satisfy some of the linear equations in \eqref{eq:NoInterferenceNullSpaceCond}. Note that this solution is optimal when $d_a=1$. Moreover, although extending the alternating minimization solution to $K_s>1$ in not trivial, interference leakage minimization and DOF-preserving solutions can be adopted for arbitrary $K_s$. 

\section{Numerical Results} \label{sec:NumericalResults}
\indent Consider an existing $3$-user $2\times 2$ MIMO IA network where, through IA \cite[Section III.A]{Peters2011}, each transmitter sends a single stream to its corresponding receiver. For the rest of this section, we add nodes to this network and compare the total achievable sum rate of the resulting new network. All the channel and additive noise coefficients are distributed as $\mathcal{CN}(0,1)$.

\begin{figure}[t]
 \centering
\includegraphics[width=3.2in]{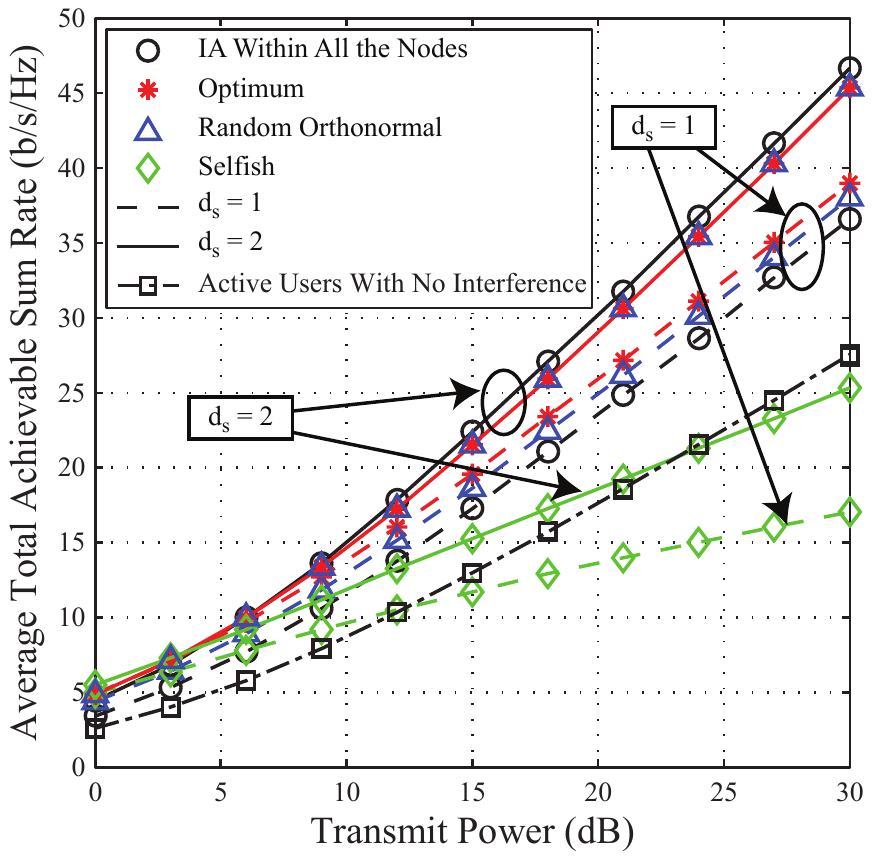}
\caption{Average total achievable sum rate versus transmit power for varying precoder designs for the single $5\times 5$ secondary node arriving to an already established $3$-user $2\times 2$ MIMO IA network. As expected, ignoring the active users results in the worst performance and reduces the attained multiplexing gain. Restricting the interference from the secondary transmitters at the active receivers increases the multiplexing gain of the network and can achieve the same DOF as if all the nodes had done IA together. Moreover, random orthonormal precoding in the interference subspace of the active receivers achieves an acceptable performance close to the optimum design.}
\label{fig:OptPrecoderNoInterferenceSingleNU}
\end{figure}

\textbf{Single secondary user not interfering with the active users} Assume $K_s=1$, $M_s=N_s=5$, and $d_s=1,2$. The secondary user precoder can be designed using the \textit{optimum} method presented in Lemma \ref{Lem:ZeroInterferenceSumRateMaximizingCognitivePrecoding} or the \textit{random orthonormal} and the \textit{selfish} designs discussed in Section \ref{subsec:SingleNewUserWithEnoughAntennas}. The average achievable sum rate of this network for various secondary user precoders versus $\gamma_o$ is shown in Fig. \ref{fig:OptPrecoderNoInterferenceSingleNU}. For reference we show the sum rate when all the nodes cooperate through IA. As expected, the selfish precoding has the worst performance where, due to the uncoordinated interference, the DOF of the whole network is less than the original IA network. Moreover, both the optimum and random orthonormal precoders achieve $5$ DOF, the maximum DOF attainable through IA (for $\hbox{DOF}>5$ algorithm of \cite[Section III.A]{Peters2011} does not converge). At low $\gamma_o$, however, where interference and noise are not distinguishable, the selfish design has the best performance. 

\begin{figure}[t]
\centering
\includegraphics[width=2.848in]{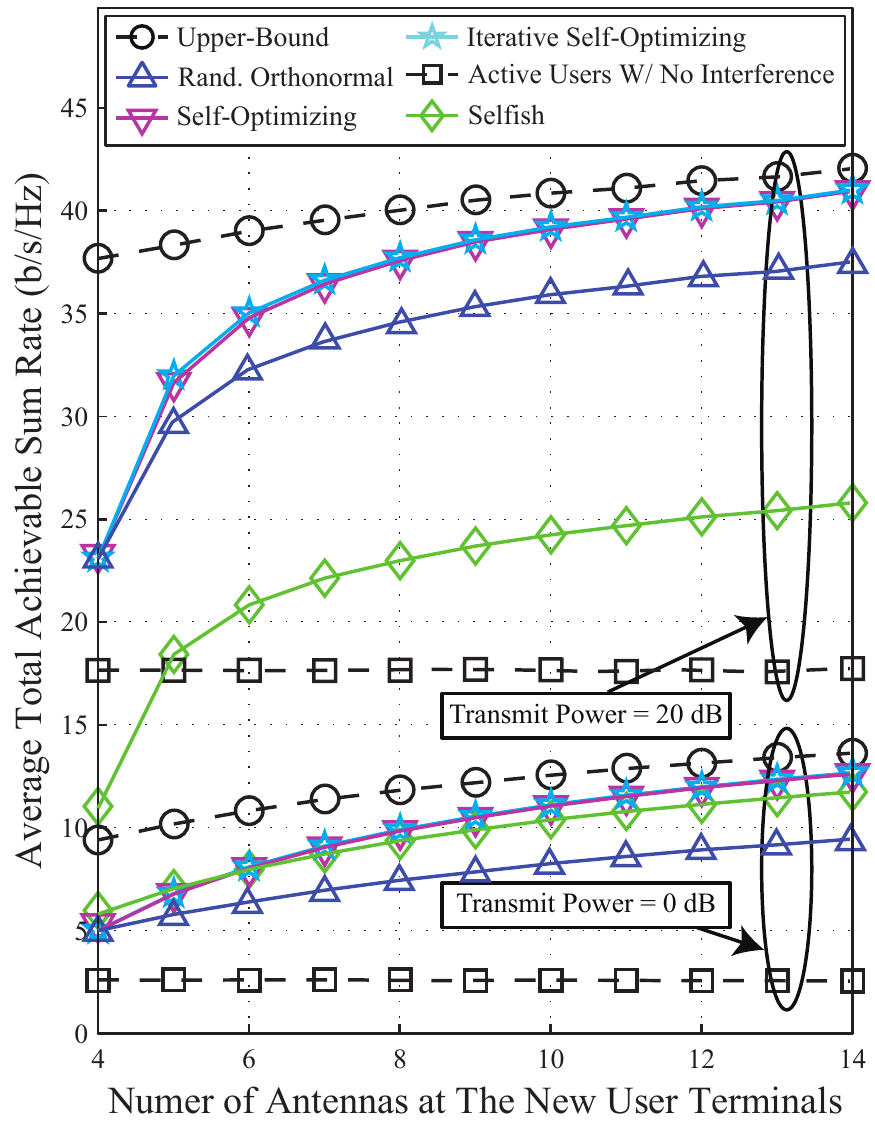}
\caption{Average total achievable sum rate versus $N_s$ at two secondary transmitters arriving to a $3$-user $2\times 2$ MIMO IA network for two SNR values. For small $M_s$, the selfish design is optimum when the network is noise limited, but in the high SNR regime, restricting the interference at the active receivers considerably increases the sum rate. As expected, when $M_s=N_s= K_a+1 = 4$, all the interference avoiding designs have the same performance. Moreover, diminishing returns of 
increasing $N_s$ is more apparent at low SNR.}
\label{fig:OptPrecoderNoInterferenceTwoNU}
\end{figure}

\textbf{Two secondary users not interfering with the active users} Now, assume two $M_s\!\times\!N_s$ MIMO pairs ($M_s\!=\!N_s$) are added to the network of active users. The secondary users can adopt \textit{selfish} precoding, \textit{random orthonormal} precoding, the \textit{self-optimizing} precoding of Corollary \ref{Col:ForgettingAboutTheSecondGuy}, or the \textit{iterative self-optimizing} precoding of Corollary \ref{Col:SuccessiveForTwoUsers}. The average total achievable sum rate versus $M_s$ for various precoders at two values of $P_a$ is shown in Fig. \ref{fig:OptPrecoderNoInterferenceTwoNU}. The depicted upper-bound is for the case of no interference among the secondary users or between the active and secondary users. At $\gamma_o\!=\!0$ dB when $N_s\!-\!\sum\limits_{i=1}^{K_a}d_i$ is small, in contrary to the high $\gamma_o$, the selfish design is outperforming the other methods indicating the available spatial dimensions are better used to combat noise than interference. As discussed before, for ${M_s\!=\!N_s\!=\!K_a\!+\!1}$, after enforcing the \textit{zero-impact} constraint on $\mathbf{F}_k$ for $k\!>\!K_a$, no further optimization of the precoders is possible explaining why  \textit{self-optimizing}, \textit{iterative self-optimizing} and \textit{random orthonormal} precodings have the same performance for $M_s\!=\!N_s\!=\!4$. Random orthonormal precoding requires considerably less overhead compared to the other methods but achieves an acceptable performance. Also, the diminishing returns of increasing $M_s$ is more sever at low $\gamma_o$. 

\begin{figure}[t] 
 \centering
\includegraphics[width=3.2in]{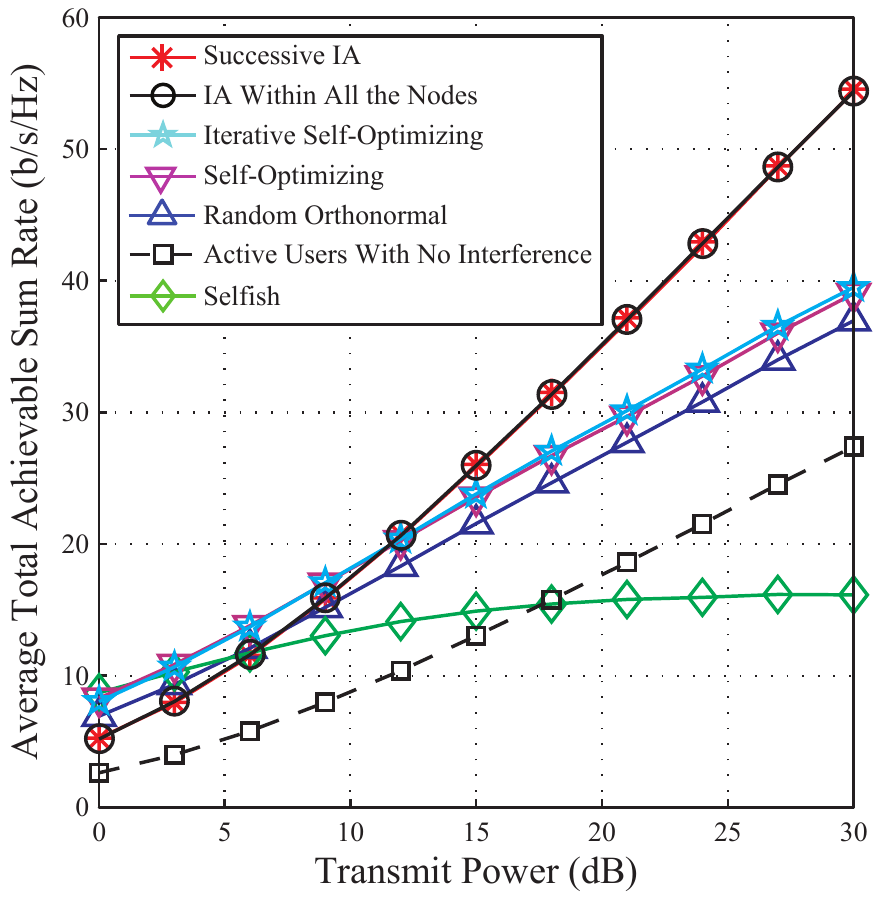}
\caption{Average total achievable sum rate versus transmit power for varying precoder designs at the three $5\times 5$ secondary transmitters arriving to a $3$-user $2\times 2$ MIMO IA network. Ignoring the interference terms, although being optimal in the low SNR (and low interference-to-noise) regime \cite{Etkin2008}, reduces the total DOF of the network. Moreover, only \textit{successive interference alignment} achieves the same DOF ($6$ in this case) as if all the nodes had jointly performed IA.}
\label{fig:OptPrecoderNoInterferenceThreeNU}
\end{figure}

\textbf{Three secondary users not interfering with the active users} For the case of $K_s=3$ and $M_s=N_s=5$, we consider \textit{selfish}, \textit{random orthonormal}, \textit{self-optimizing} design of Corollary \ref{Col:ForgettingAboutTheSecondGuy}, \textit{iterative self-optimizing} design of Corollary \ref{Col:SuccessiveForTwoUsers}, and successive IA precodings. The average total achievable sum rate is shown in Fig. \ref{fig:OptPrecoderNoInterferenceThreeNU}. In this case, as discussed in Conjecture \ref{Conj:MultiLeveLIA}, successive IA achieves the same DOF as IA over the whole network. The other methods, because of uncoordinated interference at the secondary receivers, cannot achieve more DOF than the original network of the active users. In other words, from Conjecture \ref{Conj:MultiLeveLIA}, when $M_s < (K_ad_a+K_sd_s)$ and $d_a=d_s=1$, only successive IA can possibly achieve more DOF than the original IA network. 

\begin{figure}[t]
 \centering
\includegraphics[width=3.2in]{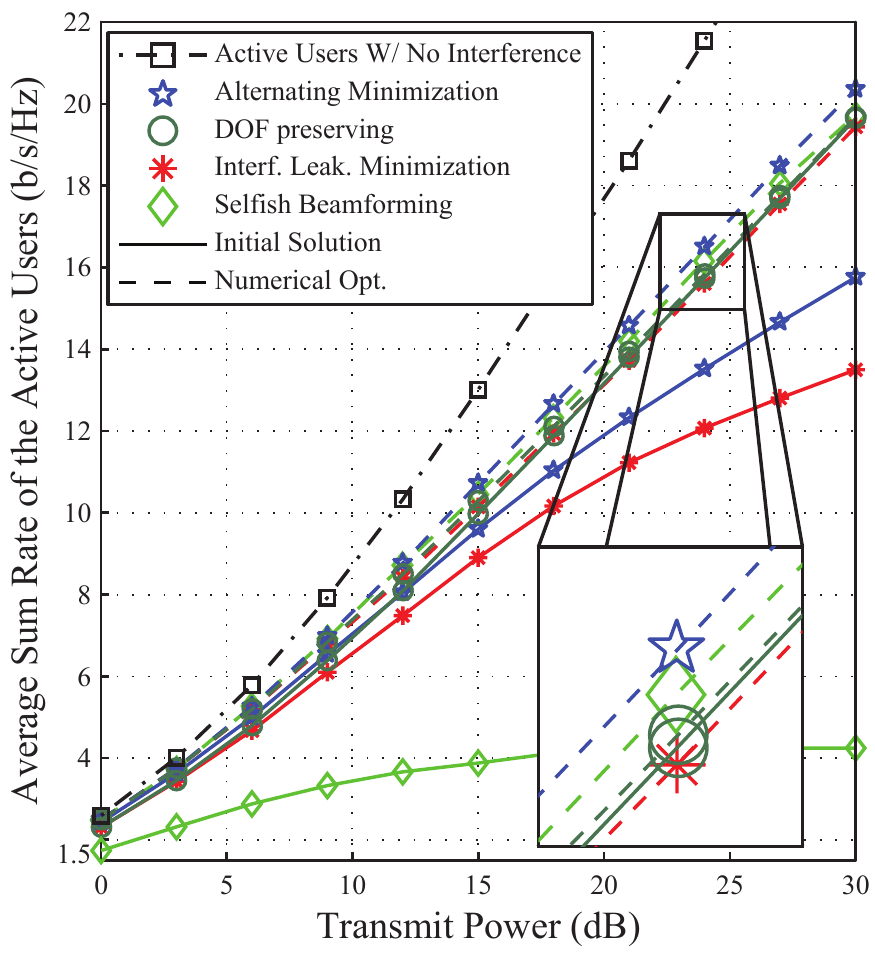}
\caption{Average total achievable sum rate versus transmit power for varying precoder designs for a single $3\times 3$ secondary user arriving to a $3$-user $2\times 2$ MIMO IA network. As expected, MGM initialized with the alternating minimization solution achieves the best performance. The DOF preserving solution outperforms the other initial solutions at high transmit power while selfish and equal gain precodings result in the worst performance.}
\label{fig:OptPrecoderWithInterferenceSingleNU}
\end{figure}

\textbf{Single secondary user interfering with the active users} Consider the case where $K_s=1$ and $M_s=N_s=3$. Here, $N_s$ does not satisfy \eqref{eq:NumberOfNUTransmitterAntennas_ZeroInterferenceToCU} and the new node is bound to degrade $R_{\rm{sum}}^{\rm{au}}$. Average $R_{\rm{sum}}^{\rm{au}}$ versus $\gamma_o$ for the MGM algorithm initialized with the three solutions provided in Section \ref{sec:SecondaryUsersInterfereingWithTheActiveUsers} is depicted in Fig. \ref{fig:OptPrecoderWithInterferenceSingleNU}. The alternating minimization algorithm approximately maximizes the sum rate of the active users and, as expected, the MGM algorithm converges to the best solution when initialized with it. In addition, the DOF preserving initial solution results in the best performance at high transmit power among all the initial solutions. Moreover, \textit{selfish} precoding can be considered as a random initialization for the MGM algorithm and attains a good performance although the initial solution itself drastically decreases $R_{\rm{sum}}^{\rm{au}}$. 

\begin{figure}[t]
 \centering
\includegraphics[width=3.2in]{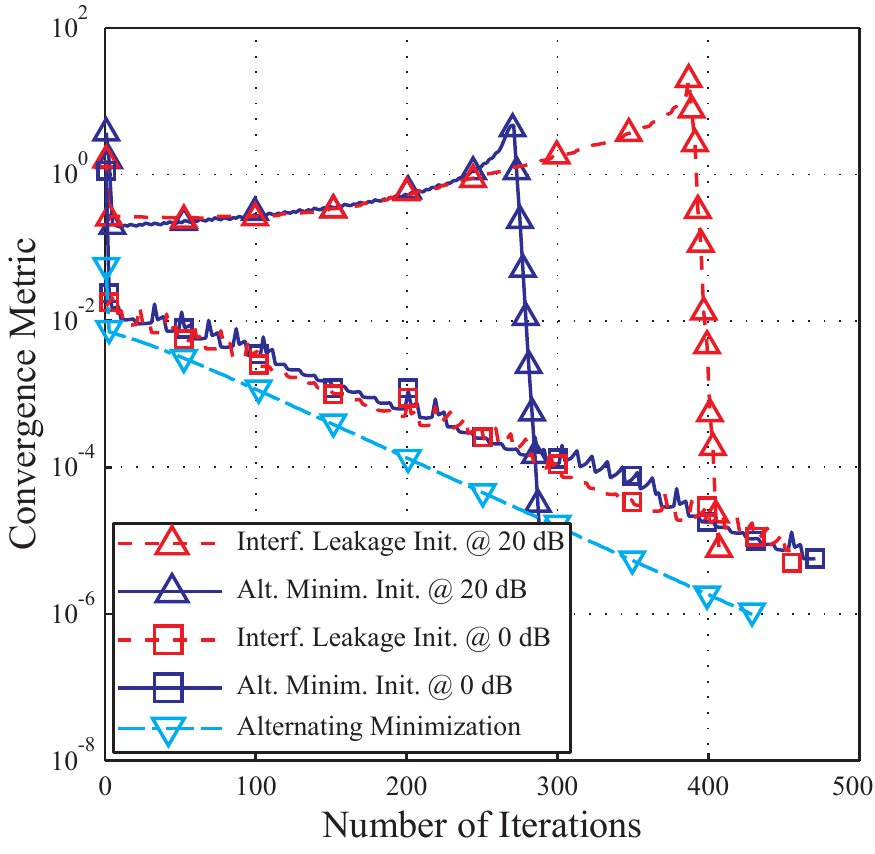}
\caption{Worst convergence rate of the MGM and the alternating minimization algorithms over $1000$ randomly generated channels. The convergence metric for the MGM algorithm is the magnitude of the descent direction on the Grassmann manifold and for the alternating minimization algorithm is the Frobenius norm of the difference between the $\mathbf{F}_s$ at two consecutive iterations. Both the algorithms converge in less than $500$ iterations in the worst case with the mean convergence rate of less than $40$.}
\label{fig:ConverganceAM_1NU}
\end{figure}

\textbf{Convergence of the MGM and the alternating minimization algorithms} Both the MGM and the alternating minimization algorithms are guaranteed to converge to a (sub) optimum solution (see \cite[Section VII.A]{Manton2002} and Section \ref{subsec:AM}). The convergence metric for the MGM algorithm is the magnitude of the descent direction on the Grassmann manifold and for the alternating minimization algorithms is the Frobenius norm of the change in $\mathbf{F}_s$ at two consecutive iterations. The worst convergence rate of the MGM algorithm at two values of $\gamma_o$ initialized with the solutions of Section \ref{subsec:AM} and \ref{subsec:IL_Min} together with the slowest convergence rate of the alternating minimization algorithm over $1000$ randomly generated channels are shown in Fig. \ref{fig:ConverganceAM_1NU}. Note that the alternating minimization algorithm does not depend on $\gamma_o$. Both algorithms converge in less than $500$ iterations (the mean number of iterations of all the cases was less than $40$).

\section{Conclusions} \label{sec:Conclusion}
In this paper, we derived sum rate and DOF optimum/suboptimum precoders for a set of secondary users arriving to an already established network of active users performing MIMO IA. We showed how knowledge of the interference subspaces at the active receivers can be used to mitigate the performance degradation caused by the secondary users and how, in some networks, gains of IA are obtainable even if all the users do not perform IA together. We derived the minimum number of secondary transmit antennas, the \textit{zero-impact} threshold, required for using the network resources without degrading the sum rate of the active nodes.  

\indent When the secondary users have more than the \textit{zero-impact} antennas, optimum and suboptimum precoders maximizing the achievable sum rate of the secondary users and the total DOF of the network were found for up to two and more than two secondary users, respectively. For up to two secondary users, we showed that a total achievable sum rate close to the upper-bound is obtainable. For more than two secondary users, the proposed successive IA techniques, for certain (and plausible) combination of number of secondary users and antennas, achieves the same DOF as if all the users had simultaneously performed IA. The successive IA method, as a decentralized IA technique, can be adopted in next generation wireless communication systems where different categories of nodes, such as base stations, femtocells and user terminals with different number of antennas and different levels of cooperation have to coexist. 

\indent When the secondary users have less than the \textit{zero-impact} antennas, we proposed a precoder design based on steepest descent search on the Grassmann manifold. We derived three initial solutions for the numerical search, maximizing the sum rate of the active users at high SINR, minimizing the interference power \textit{leakage} in a least squares sense, and maximizing the obtainable DOF in the IA network. The sum rate maximizing initial solution was the most complex and resulted in the highest active users' sum rate. 

\indent We assumed global CSI without considering the associated overhead. Also, with increasing number of antennas the interference mitigation power of the secondary nodes increases, but so do the overhead associated with training the additional links and relaying their CSI. Quantifying such overhead costs and the effects of a limited capacity feedback link is left for future work.  

\bibliographystyle{IEEEtran}
\bibliography{../../Bib/IEEEabrv,../../Bib/IA_main_ver2,../../Bib/IA_UserAdmission}

\clearpage

\end{document}